\documentclass[twocolumn,showpacs,preprintnumbers,amsmath,amssymb,pre]{revtex4-1}
\usepackage{graphicx}
\usepackage{dcolumn}
\usepackage{overpic}
\usepackage{float} 

\begin{document}

\title{Random-walk topological transition revealed via electron counting}

\author{ G. Engelhardt$^{1,2}$    }

\author{ M. Benito$^{3,4}$}

\author{ G. Platero$^3$}

\author{ G. Schaller$^1$}

\author{ T. Brandes$^1$    }

\affiliation{%
$^1$Institut f\"ur Theoretische Physik, Technische Universit\"at Berlin, Hardenbergstr. 36, 10623 Berlin, Germany \\
$^2$Beijing Computational Science Research Center, Beijing 100048, P. R. China,
$^3$Instituto de Ciencia de Material de Madrid, CSIC, 28049 Madrid, Spain\\
$^4$Department of Physics, University of Konstanz, D-78457 Konstanz, Germany
}



\begin{abstract}
	The appearance of topological  effects in systems exhibiting a non-trivial topological band structure strongly relies  on the coherent wave nature of the equations of motion. Here, we reveal topological dynamics in a classical stochastic random walk version of the Su-Schrieffer-Heeger  model with no relation to coherent wave dynamics. We explain that the commonly used topological invariant in the momentum space translates into an invariant in a counting-field space. This invariant gives rise to clear signatures of the topological phase in an associated escape time distribution.
\end{abstract}
\maketitle

\textit{Introduction.}

Starting from topological insulators and superconductors~\cite{Hasan2010,Bernevig2013,Chiu2016}, the manifestation of topological band structures has been investigated in various contexts. Relying on the wave nature of the dynamics, topological band structure effects also appear in bosonic and classical systems~\cite{Engelhardt2015,Engelhardt2016,Engelhardt2017,Peano2016,Suesstrunk2015,Poli2015,Lee2017,Bello2016}. 
Furthermore, topological effects are manifested in quantum walk problems~\cite{Ramasesh2017,Flurin2017,Preiss2015,Kempe2003,Asboth2013,Asboth2012,Kitagawa2010,Kitagawa2012}. Thereby, a quantum mechanical particle moves randomly on a lattice, where the movement is determined by a quantum mechanical equation of motion. For this reason, the dynamics of the quantum  walk inherits the wave nature of quantum mechanics which consequently can give rise to  topological band structure effects.

\begin{figure}[t]
\includegraphics[width=0.9\linewidth]{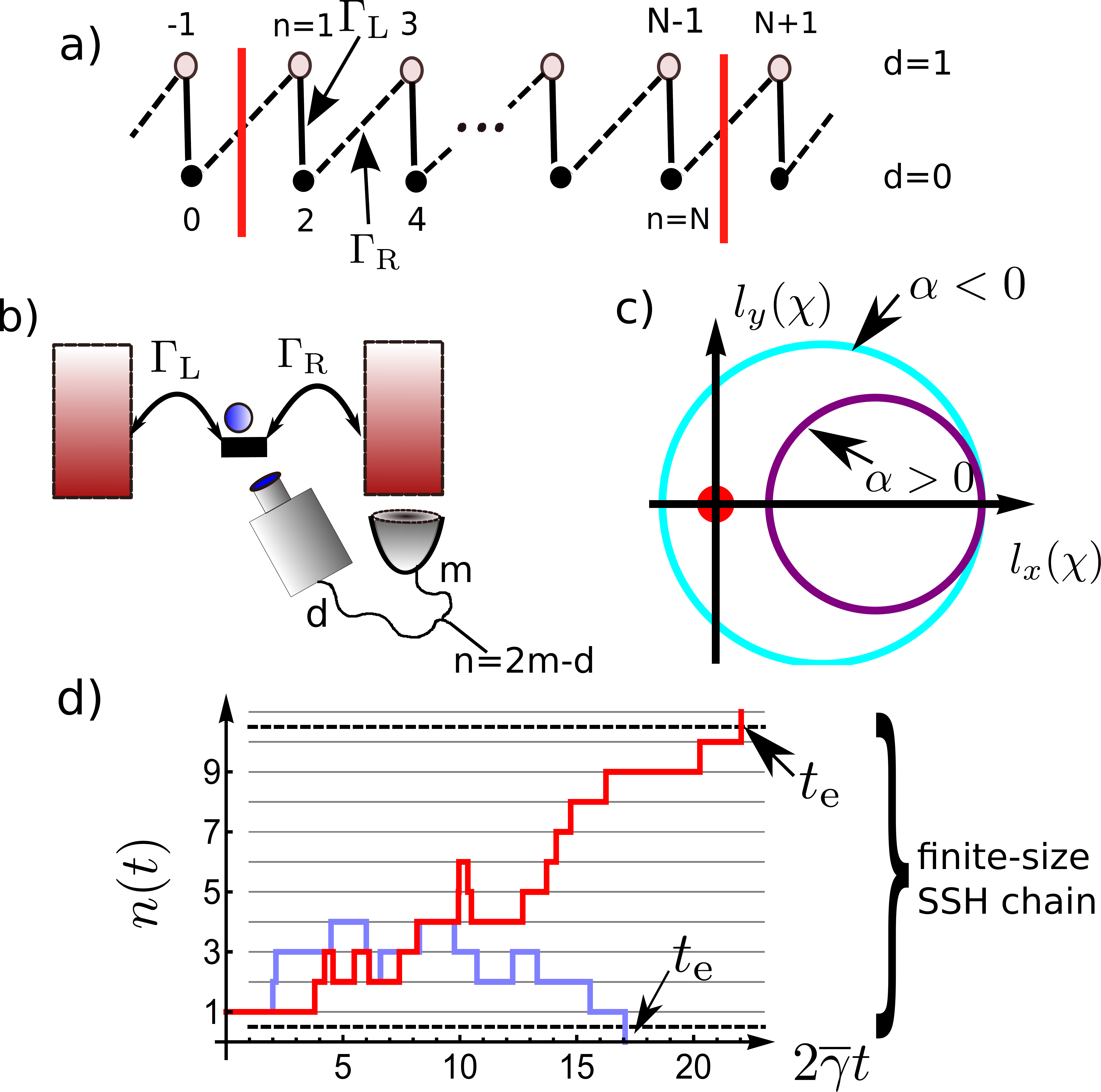}
\caption{(a) Stochastic random walk on a lattice with SSH coupling geometry. The state of the system $\left|n\right>$ changes randomly as a function of time with alternating hopping probabilities according to Eq.~\eqref{eq:dynamicRandomWalk}. (b) This random walk describes the dynamics of a SET connected to two Markovian leads with properly adjusted chemical potentials, which are coupled to the dot with strengths $\Gamma_{\rm R}$ and $\Gamma_{\rm L}$. By counting the number of particles tunneled into the right reservoir $m$ and monitoring the dot occupation $d\in \{0,1\}$, one can infer the state $n=2 m- d$.
(c) Illustration of the  TI in the SSH model, which is equivalent to a winding of the curve $ \left[l_{x}(\chi),l_y(\chi)\right]$ (depending on $\Gamma_{\rm R/L}$) around the origin.
(d)  Stochastic trajectory of $\left| n\right>_t$ as a function of time in units of $2\overline \gamma=\Gamma_{\rm R}+\Gamma_{\rm L}$. At time $t_{\rm e}$   the state   escapes from the region $\left\lbrace 1,...,N \right\rbrace$, resembling the quantum SSH model with an open boundary condition.
}
\label{fig:overview}
\end{figure}

In this article, we abandon the requirement of a wave-like motion and consider a purely  stochastic random walk in a classical fashion. As we explain in the following, a properly designed system still exhibits clear features of a topological coupling geometry. We choose a random walk version of the celebrated Su-Schrieffer-Heeger (SSH) model to explain this effect. A sketch of the system is depicted in Fig.~\ref{fig:overview}(a).  The SSH model consists of a linear chain of nodes with staggered coupling strength and is presumably the simplest model exhibiting topological effects~\cite{Su1979,Asboth2016,Gomez-Leon2013}. In our investigation, the state of the system can hop randomly   along the SSH chain.

\begin{figure*}
\centerline{\includegraphics[width=0.9\linewidth]{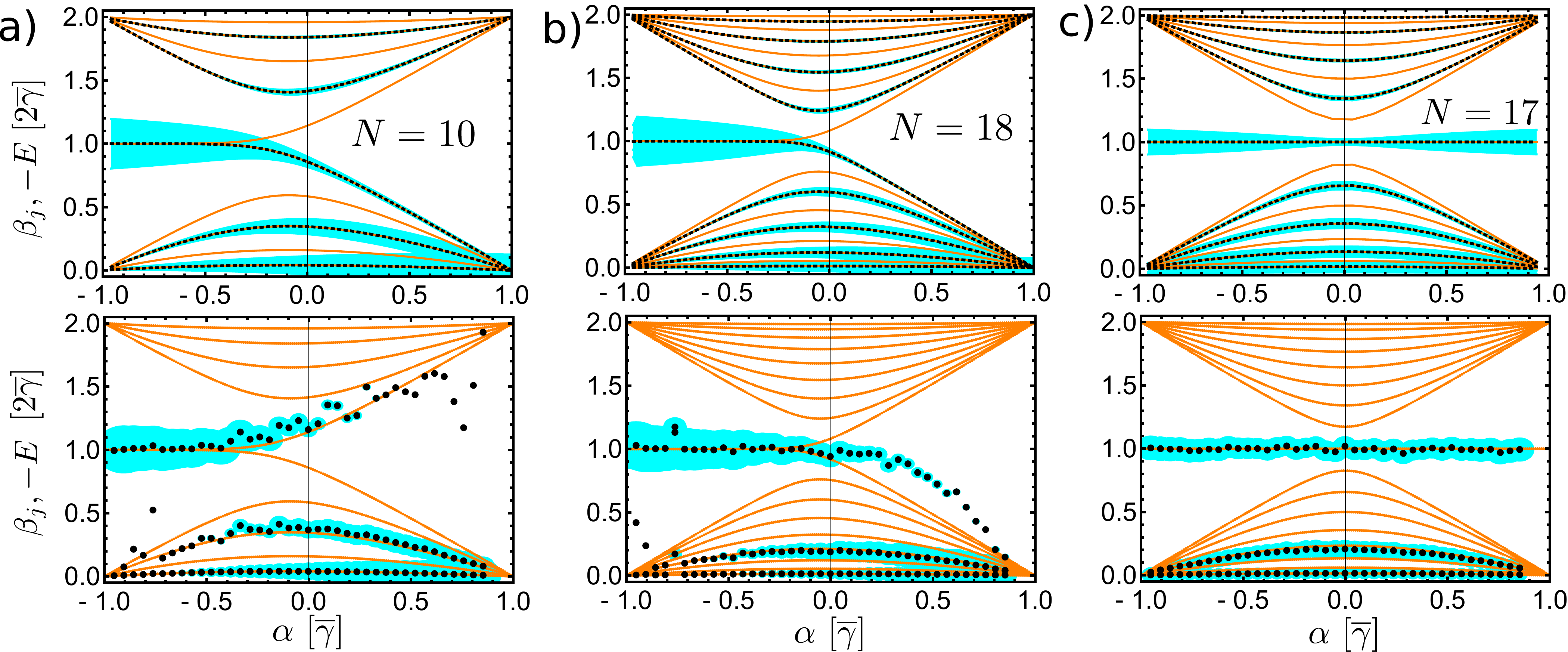}}
\caption{
Comparison of the spectrum  of $\mathbf L_{\rm SSH}$ in Eq.~\eqref{eq:finitSize} with  open boundary conditions and the exponent spectrum $\beta_i$ of the integrated ETD according to Eq.~\eqref{eq:integratedProbDistribtution}. In  (a), (b) and (c) we depict the results for  the   chain lengths $N=10$, $N=18$ and  $N=17$, respectively.
The orange (solid) lines depict the spectrum of the matrix $\mathbf{L}_{\rm SSH}$.  
The top panels depict the exact results obtained by  a diagonalization. The dotted exponent spectrum depicts the levels with non-vanishing coefficients $A_j$ (see Eq.~\eqref{eq:integratedProbDistribtution}). The coefficients $A_j$ are represented by the blue regions, whose width is proportional to  $A_j$.  The bottom panels depict the exponent spectrum obtained by a simulation of random trajectories. To construct the integrated ETD we used $i_{\rm max}=10^5$ random trajectories, which have been fitted  with $K=3$.
}
\label{fig:expSpectrum}
\end{figure*}%

Due to the underlying topological coupling geometry, one can define a topological invariant (TI) based on the generalized density matrix, where the \textit{counting field } takes the role of momentum in common topological band structures.
We show that a properly defined escape time statistics  will reveal the topology. Thereby, the SSH model with an open boundary condition is associated to the escape time from a finite region of the SSH random walk as depicted in Fig.~\ref{fig:overview}(d). 

Our approach requires a detailed counting statistics with a large number of experimental runs. In order to obtain the required  amount of data, we suggest to implement the random walk using a single-electron transistor (SET), which in the full-counting space is described by a SSH random walk.

 Understanding  the relaxation dynamics of mesoscopic devices is of fundamental interest in the development of  mesoscopic electronic devices~\cite{Schulenborg2016,Schulenborg2014} as single-electron emitters~\cite{Feve2007,Blumenthal2007}, quantum pumps~\cite{,Buitelaar2008,Giazotto2011,Splettstoesser2005}, or solid state qubits~\cite{Hanson2003}. A detailed counting statistics can provide  information about the underlying processes and correlations arising in  mesoscopic devices, as universal oscillations investigated in Refs.~\cite{Flindt2009,Ptaszyfmmodecutenlseniski2017}. A basic theoretical knowledge is required to develop  schemes to control the counting statistics~\cite{Wagner2017,Brandes2017}. In this regard, our findings contribute to the fundamental understanding of  processes being active in such kind of systems.
The possibility of including feedback operations allows to study even more sophisticated models~\cite{Supplementals}.

\textit{The system.}
We consider a classical random walk on a one-dimensional lattice (Fig.~\ref{fig:overview}(a)). The sites are labeled by $n=0,\pm1,\pm2,\dots$.  The random walk of the state $\left|n\right>_t$ from time $t$ to time $t+d t$ is determined by the transition probabilities $p_+, p_- , p_s$ which are defined by
\begin{equation}
\left|n\right>_t\rightarrow 
\begin{cases}
\left|n+1\right>_{t+d t} &\text{with} \quad p_+=\left (\overline \gamma -(-1)^n \alpha\right)d t   \\
\left|n-1\right>_{t+d t} & \text{with}\quad p_-=\left (\overline \gamma +(-1)^n \alpha\right)d t  \\
\left|n\right>_{t+d t} &  \text{with}\quad  p_s =1- p_+ -p_-,
\label{eq:dynamicRandomWalk}
\end{cases}
\end{equation}
where $2 \overline \gamma dt$ is the probability that the system escapes from site $n$ within an infinitesimal time step $ d t$. We  impose a coupling geometry with an alternating hopping probability. The parameter $\alpha$ determines a jump bias so that for $\alpha\neq 0$ a jump to either $\left|n+1\right>$ or $\left|n-1\right>$ is preferred. The coupling geometry is thus analog to  the SSH model~\cite{Su1979,Asboth2016}. Eq.~\eqref{eq:dynamicRandomWalk} appears in the transport   dynamics of a SET, i.e., a quantum dot connected to two electronic leads,  when the chemical potentials match the on-site energy of the quantum dot. In this case, the effective coupling parameters are $\Gamma_{\text L/\text R}= \overline \gamma \pm\alpha$   (see Fig.~\ref{fig:overview}(b)~\cite{Supplementals,Bonet2002}). 

For the following analysis, we  introduce the parametrization $n=2m -d$ with integer $m$ and $d\in \left\lbrace0,1 \right\rbrace$.
The  probability distribution corresponding to Eq.~\eqref{eq:dynamicRandomWalk} follows the equation 
\begin{equation}
	\frac{d}{dt }\underline \rho_m = \mathbf L_0 \underline \rho_m  + \mathbf L_+\underline \rho_{m-1} +\mathbf L_- \underline \rho_{m+1},
	\label{eq:EomProbabilities}
\end{equation}
where $\underline \rho_m= \left( p_{m,1},p_{m,0} \right)^{\text T}$ contains the probabilities $ p_{m,d}=p_{n=2m-d}$ that the system is in state $\left|n=2m -d \right>$ and
\begin{equation}
\mathbf L_0 = 
\left(
\begin{array}{cc}
-2 \overline \gamma & \Gamma_{\text L} \\
   \Gamma_{\text L} & -2 \overline \gamma
\end{array}
\right),
\,
\mathbf L_+ = 
\left(
\begin{array}{cc}
0 & \Gamma_{\text R} \\
  0  & 0
\end{array}
\right),
\,
\mathbf L_- = \left(\mathbf L_+\right)^{\rm T}
.\nonumber
\end{equation}
Regarding the SET (Fig.~\ref{fig:overview}(b)), $\mathbf L_+$ describes a jump of an electron from the  right reservoir into the dot, while the non-diagonal entries of  $\mathbf L_0$ describe the jumps related to the left reservoir. Additionally, the diagonal terms $-2 \overline \gamma=-\Gamma_{\rm L}-\Gamma_{\rm R} $ are responsible for the correct normalization of the probability distribution $\sum_n p_n=1$. The index $m$ can thus be interpreted as the number of particles having jumped out of the right reservoir. For instance, if the initial state is  $p_n(0)=\delta_{n,1}$, then we have one particle $m=1$  tunneled out of the right reservoir and a dot occupation of $d=1$.

\textit{Relation to the  Schr\"odinger equation and topology.}
By replacing $d/dt \rightarrow {\rm i}d/ d t$, Eq.~\eqref{eq:EomProbabilities} becomes equivalent to the Schr\"odinger equation of a particle in the quantum mechanical SSH model, when interpreting $\underline\rho_m$ as the corresponding wave function. 
We emphasize that the introduction of the complex unit $\rm i$  is more than  a reparametrization, but renders one real-valued equation into a complex-valued (thus, two  real-valued) equation(s). In consequence, we obtain the wave-like  Schr\"odinger equation, so that we can expect different kinds of physical dynamics.

Yet, the formal analogy to the  quantum SSH Hamiltonian gives rise to  topological effects in a stochastic random walk. To see this, we  apply the concepts known from the quantum  model to introduce a TI.

 By applying a Fourier transformation  Eq.~\eqref{eq:EomProbabilities} becomes
\begin{align}
	\underline {\dot \rho}(t,\chi) &= \mathbf L_{\chi }\, \underline \rho(t,\chi), \qquad
	\mathbf L_{\chi } =- 2 \overline \gamma \mathbf 1 + \underline  l (\chi ) \cdot \underline{\boldsymbol \sigma} ,
	%
	\label{eq:EoMcountingField}
 \end{align}
where $\underline l(\chi)=\left[l_{\rm x}(\chi ),l_{\rm y}(\chi )\right]$ with $l_{\rm x}(\chi )  = \Gamma_{\text L} + \Gamma_{\text R} \cos(\chi) $, $l_{\rm y}(\chi )  = \Gamma_{\text R} \sin(\chi) $ and $\underline{\boldsymbol \sigma}=\left[\boldsymbol \sigma_{\rm x},\boldsymbol \sigma_{\rm y} \right]$ with $\boldsymbol \sigma_{\rm x,y,z}$ denote the usual Pauli matrices. 
Importantly, $ (1,1)\cdot \underline \rho(t,\chi)$ represents the moment generating function, whose derivatives with respect to $\chi$ are the moments of the probability distribution $p_m(t)$ that the system is in either the state $n=2m$ or $n=2 m-1$.

 The matrix $\mathbf  L_\chi$ is  equivalent to the matrix representation of the SSH Hamiltonian in momentum space when identifying the counting field $\chi$ with the momentum. In particular, we can define two topological phases for $\alpha>0$ (trivial) and $\alpha<0 $ (non-trivial) which are characterized by a TI: $2 \pi W = \int_{-\pi}^{\pi}d\chi \frac{d}{d\chi} \arg\left[l_{\text y}(\chi)/ l_{\text x} (\chi) \right] $, which is $W=0$ (trivial) or $W=1$ (non-trivial). This invariant is equal to the winding of the curve $\underline l(\chi)$ around the origin as illustrated in Fig.~\ref{fig:overview}(c). We note that $W$ is also directly linked to the geometrical Berry (or Zak) phase $ \phi_{\rm Berry}=W/2$~\cite{Berry1984}. Importantly, the definition of an invariant requires that there is no term proportional to $\boldsymbol \sigma_z$ appearing in Eq.~\eqref{eq:EoMcountingField}. This is guaranteed by the existence of a \textit{ chiral} symmetry in the equations of motion ~\cite{Asboth2016,Chiu2016}. For our system Eq.~\eqref{eq:dynamicRandomWalk} this means that the probability to escape from the even and odd sites $n$ is equal~\cite{noteChiral}.

The strict quantization of  $W$ in an infinite-size system  has a strict consequence for the finite-size (quantum) SSH Hamiltonian defined on the  sites $n \in \left\lbrace1,\dots,N \right\rbrace$  with an open boundary condition~\cite{Asboth2016}, i.e., with zero coupling between  $n=0$, $n=1$  and  $n=N$, $n=N+1$. 
The corresponding spectrum   exhibits topologically protected \textit{midgap modes} if the system is in the non-trivial phase. We  depict such spectra in Fig.~\ref{fig:expSpectrum} with orange solid lines for different chain lengths $N$. The symmetry around $E=- 2\overline \gamma $ of the spectrum is a consequence of the chiral symmetry. In Figs.~\ref{fig:expSpectrum}(a) and (b) we depict the spectrum for an even number of sites $N$.  We find a pair of  energies for $\alpha>0$ at
 the inner boundaries of the bands which merge for decreasing $\alpha$ and  become degenerate for $\alpha\lesssim 0$. This is a typical signature in the non-trivial phase of the SSH model. Due to the inversion symmetry in the SSH chain for $N$ even, the wave function of these two midgap states at $E=- 2\overline \gamma$ are symmetric and antisymmetric upon inversion, respectively~\cite{Asboth2016}. 
For  $N$  odd the chain also exhibits a generalized inversion symmetry~\cite{Supplementals}. The corresponding spectrum is depicted in Fig.~\ref{fig:expSpectrum}(c). We observe for all $\alpha$ a midgap state, whose wave function is localized close to $n=1$  ($n=N$) for $\alpha<0$ ($\alpha>0$).

\textit{Escape time distribution.}
The TI described by $W$  is a theoretical classification of the topological phase which can be hardly determined in   experiment. However, the close analogy to the SSH model and the localized midgap modes allow for a different detection scheme.

To this end, we use the existence or absence of midgap modes in a finite-size system. We construct an associated escape time distribution (ETD), which  resembles  an open boundary condition of the quantum SSH model: we divide the originally infinite chain in Fig.~\ref{fig:overview}(a) in three parts.
The middle section consisting of sites $n \in  \left\lbrace1,\dots,N \right\rbrace$ constitutes the random walk analog of the SSH model with an open boundary condition:
 Defining the probability vector $\underline \rho= \left(p_{n=1},\dots,p_{n=N} \right)$ containing the probabilities of the middle section, we can represent  Eq.~\eqref{eq:EomProbabilities} as
\begin{equation}
	\underline{\dot \rho} = \mathbf L_{\text{SSH}}\underline \rho + \underline J_{1} p_{0}+ \underline J_{N} p_{N+1},
	\label{eq:finitSize}
\end{equation}
where the entries of the \textit{jump vectors} read $\left(\underline J_{1} \right)_{k}= \Gamma_{\text R} \delta_{1,k}$ and $\left(\underline J_{N} \right)_{k}= \left[\overline \gamma-(-1)^N\alpha \right] \delta_{N,k}$.
Importantly, $\mathbf L_{\text{SSH}}$ is equivalent to the quantum  SSH Hamiltonian with an open boundary condition.
We investigate  the time $ t_{\text e}$ at which  the state  escapes from the finite-size SSH section  when initiated at a SSH site at $t_0=0$. This means that the experimentalist creates the open boundary condition by stopping  the experimental run when the state leaves the finite-size SSH section which is feasible with current experimental technologies~\cite{Wagner2017}.
 \begin{figure}[t]
 \includegraphics[width=1.\linewidth]{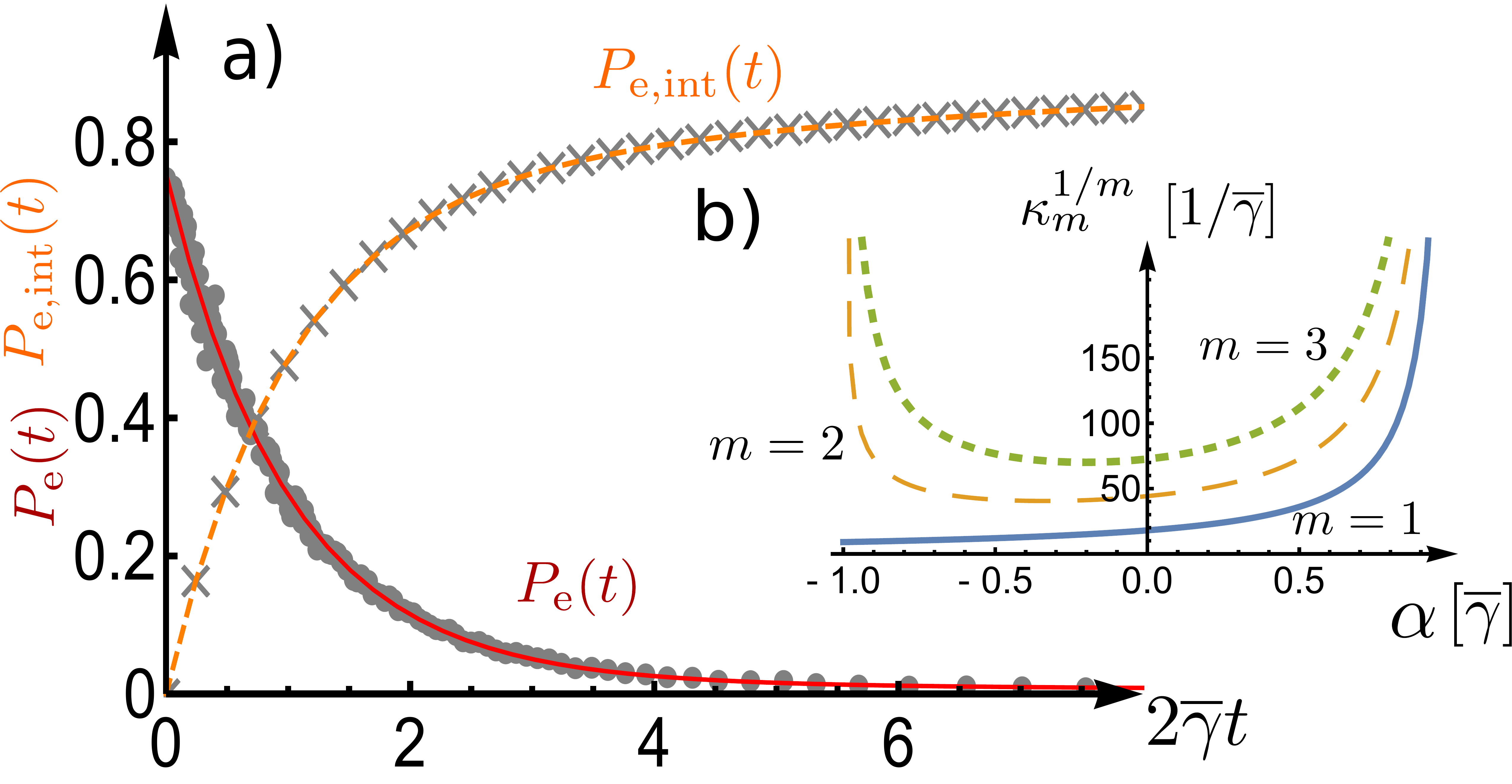}
 \caption{(a) depicts the reconstructed escape time distribution (ETD) $P_{\rm e} (t)$(dots), the reconstructed integrated ETD $P_{\rm e,int} (t)$ (crosses) and  the corresponding fitted curves with  solid and dashed lines for $\alpha =-0.5\overline \gamma$. (b) Cumulants $\kappa_m$ of the ETD for $m=1$ (solid), $m=2$ (dashed) and $m=3$ (dotted) as a function of $\alpha$.  The depicted  and higher cumulants do not reveal any signature of topology as, e.g., a non-analyticity at $\alpha=0$.
 }
 \label{fig:waitingTimeDistribution}
 \end{figure}%

The probability that the system escapes from the SSH chain at time $t_{\text{e}}$ reads~\cite{Brandes2008,Supplementals}
\begin{align}
	P_{\text{e}}(t_{\text{e}})&=  \underline J ^T e^{\mathbf L_{\text{SSH}} t_{\text{e}} }\underline{\rho}(0)
	             = \sum_{j=1}^{j=K} a_j e^{-\beta_j t_{\text{e}}}>0,
	             \label{eq:watingTimeDis}
\end{align}
where $\underline J^{\rm T}$ is the transpose of $\underline J = \underline J _1 +  \underline J _{N}  $.  
 For the second equality we have used the eigenvalues $E_j=-\beta_j$ and eigenstates $\underline v_j$ of $ \mathbf L_{\text{SSH}}$. The coefficients read $a_j= \left( \underline J^{\rm T} \cdot \underline v_j \right) \left( \underline v_j^{\rm T} \cdot \underline \rho(0)\right)$ and $K=N$. The time dependence of the ETD is thus determined by the eigenstates and eigenvalues of the finite-size SSH model, and consequently, of the underlying topology. 
The integrated ETD 
\begin{align}
  P_{\text{int}}(t ) = \int_{0}^{t}  P_{\textbf{e}}(t' ) dt'= 1-\sum_{j=1}^{j=K} A_j e^{-\beta_j t}
  \label{eq:integratedProbDistribtution}
\end{align}
fulfills $ P_{\text{int}}(\infty  )=1$. From Eq.~\eqref{eq:integratedProbDistribtution} we  find that $\sum_j A_j =1$ and $A_j = - a_j/\beta_j$.

In the following, we choose a symmetric initial state $p_n(t=0) = \delta_{n,1}/2+ \delta_{n,N}/2  $. An example of the resulting  ETD is depicted in Fig.~\ref{fig:waitingTimeDistribution}(a) with  a solid line, which shows its decaying character. 
Even though there is an underlying but complex relation between the exponent spectrum and the cumulants~\cite{Stegmann2017}, the moments $\mu_m = \int_{0}^{\infty}t^m P_{\textbf{e}}(t)dt $ and the associated cumulants $\kappa_m$~\cite{Schaller2014} depicted in Fig.~\ref{fig:waitingTimeDistribution}(b) do not provide direct information about the topology.

 For this reason, we continue to investigate the exponents $\beta_j$ and coefficients $A_j$ determining the integrated ETD. These are depicted in the top row of Fig.~\ref{fig:expSpectrum}. The  $\beta_j$ for  non-vanishing $A_j$  are depicted with black (dotted) lines and their coefficients $A_j$ are represented by the blue regions, whose width is proportional to $A_j$. Importantly, in Fig.~\ref{fig:expSpectrum} we can only find  every second $\beta_j$.  This is related to the (generalized) inversion symmetry of the system. For $N $ even,  the eigenstates $\underline v_j$ exhibit either  even ($v_{j,n}=v_{j,N+1-n}$) or  odd ($v_{j,n}=- v_{j,N+1-n}$) parity. Therefore, the  coefficients $a_j$ and $A_j$ for the odd eigenstates vanish as $\underline J$ has  even parity, $\left(\underline J \right)_{n}=\left(\underline J \right)_{N+1-n}$.
 A similar reasoning apply for $N$ odd~\cite{Supplementals}.
 Remarkably, the coefficients of the midgap modes for $N=10$ and $N=18$ are very similar.
 This is a consequence of the fact that the midgap  eigenvectors  only slightly depend on the system size. 
  Due to the symmetric initial condition $\underline \rho(0)$, we find also symmetric coefficients $A_j$ with respect to  $\alpha\rightarrow-\alpha $.

\textit{Detection of the topological phase.}
After investigating the dynamics on the probability level, we now return to the   random walk according to Eq.~\eqref{eq:dynamicRandomWalk}. We can reconstruct the ETD by initializing the system on a site $n$ and measuring the escape time $t_{\text{e}}$ (Fig.~\ref{fig:overview}(d)).
By repeatedly conducting this experiment and determining the escape times $t_{\text{e},i}$, with $i=1,\dots,i_{\rm max}$ we can construct the ETD and the integrated ETD~\cite{,Supplementals}. To resemble the initial state $p_n(t=0) = \delta_{n,1}/2+ \delta_{n,N}/2  $, we start half of the random trajectories on site $n=1$ and the other half on site $n=N$.
 In Fig.~\ref{fig:waitingTimeDistribution}(a), we depict  the reconstructed distributions for $\alpha =-0.5\overline \gamma$ by using $i_{\rm max}=10^5$ random trajectories.

 Fitting the reconstructed ETD with the ansatz Eq.~\eqref{eq:watingTimeDis}  provides information about  the eigenvalues  of $\mathbf L_{\text{SSH}} $.  We use the integrated ETD and Eq.~\eqref{eq:integratedProbDistribtution} instead of the ETD as this provides a higher degree of reliance for the fit parameters, in particular  for small $\beta_j$.
 We find that in Eq.~\eqref{eq:integratedProbDistribtution}  $K=3$ is sufficient to resemble the reconstructed integrated ETD with a high accuracy. The case  $K>3$ is discussed in~\cite{Supplementals}.
 In the bottom row of Fig.~\ref{fig:expSpectrum}, we depict the exponent spectrum $\left\lbrace \beta_j\right\rbrace$   obtained with this procedure.

For a short chain length  $N=10$, the   exponent spectrum agrees well with the spectrum of  $\mathbf L_{\text{SSH}}$ for $\beta_j\lesssim 2 \overline \gamma$. In particular, the midgap state with $\beta_j=1$ is clearly visible  in the  nontrivial phase for $\alpha<0$. The eigenstates with $\beta_j>2 \overline \gamma $ are not resembled by this procedure. This is a consequence of the corresponding small $A_j$ in the expansion Eq.~\eqref{eq:integratedProbDistribtution} and the fast transition dynamics related to the relative large $\beta_j$. Remarkably,  the fitting procedure resembles every second eigenvalue for $\beta_j<2 \overline\gamma$. Thus, we observe the eigenvalues  with even parity according to our previous explanation and according to  the top panel in Fig~\ref{fig:expSpectrum}(a). Moreover, corresponding to the theoretical prediction the fitted  $A_j$ are considerably larger for the midgap state as for the other $\beta_j$. For $\alpha>0$ we find some $\beta_j$ which do not fit to the spectrum of $\mathbf L_{\rm SSH}$. However, the corresponding $A_j$ are  small so that they do not significantly influence the fit quality.

For a longer chain with $N=18$ we observe similar features. In particular, we also recognize the midgap state. 
For  $N=17$,  the exponent spectrum  of the reconstructed integrated ETD resembles the main features of the $ \mathbf L_{\rm SSH}$ spectrum.  Due to the chosen initial condition, the coefficients $A_j$ are equal for $\alpha$ and $-\alpha$. This results in the symmetry  observed in Fig.~\ref{fig:expSpectrum}(c), where the midgap exponents are located at $\beta_j \approx 2\overline \gamma$ for all $\alpha$ values.

\textit{Conclusions.}
We showed that a classical random walk  on a lattice with SSH  coupling geometry exhibits a TI signaling the topological phase. This TI is defined by the generalized density matrix  as a function of the counting field $\chi$, which constitutes the analog  description of the system in momentum space known from the quantum  SSH model. This relation is reminiscent, but distinct from the investigations Refs.~\cite{Wang2017,Ren2010,Li2014,Ren2013,Benito2016,Sinitsyn2007} establishing also a link between counting statistics and topology.  We showed that the topological phase is revealed in the spectrum of fitted exponents of a properly designed ETD. Although the fitting procedure applied to the random data is sensitive to numerical details, we found that boundary modes are strongly pronounced in the exponent spectrum. This feature remains independent of the chain length, which confirms the underlying topological character in the stochastic dynamics. Even for moderately time-fluctuating rates, which keep the chiral symmetry  $\Gamma_{\text L}(t)+\Gamma_{\text R}(t)=\overline \gamma$, the presence or absence of the midgap mode should not be changed. Moreover, even for a next-nearest neighbor hopping (e.g., caused by missing a jump due to a finite detector time resolution), a topological classification is possible if there is still a chiral symmetry.
These exponents provide thus a characterization of the ETD different from the cumulants, which do not exhibit  direct information about the topology. 

The required experimental data can be generated using  quantum dots with an adjacent quantum point contact~\cite{Flindt2009}. This amount of data is in the order of magnitude needed to detect the topological dynamics. In order to enable a bidirectional particle counting required for our proposal, one could harness an experimental setup as in Refs.~\cite{Utsumi2010,Fujisawa2006}. There the direction of a particle jump (into the reservoirs or out from the reservoir) can be detected by a spatial bipartition of the quantum dot and an asymmetrically coupled quantum point contact.

To resemble the SSH dynamics and topological issues, we considered here  specially chosen chemical potentials. However, even for a general temperature and  voltage bias, the generalized master equation can exhibit fascinating (topological) effects such as exceptional points~\cite{Daryanoosh2016}. A similar escape time experiment could in this case reveal the underlying physical processes.  Moreover, the suggested setup can be  harnessed to create more complex random walks by means of feedback control as we discuss in Ref.~\cite{Supplementals}. 

The discovered topology in random walks is not restricted  to nanoelectronic devices as the  SET, but can appear in other kinds of random walk setups. In this respect it will be interesting to consider  extensions to  two or higher dimensional random walk lattices.

\textit{Acknowledgments}
The authors gratefully acknowledge financial support
from the DFG Grants 
No. BR 1528/9-1, No. SFB 910.
 This work was financially supported by the Spanish Ministry
through Grant No. MAT2014-58241-P, the FPI program and the National Natural Science Foundation of China (under Grant No.:U1530401).

\bibliography{topology}

\begin{thebibliography}{49}
\bibitem [{not()}]{noteChiral}%
  \BibitemOpen
  \href@noop {} {}\bibinfo {note} {In general, a Hermitian matrix $\mathbf L$
  exhibits chiral symmetry if there is a $x \in \mathbb R$ and a unitary
  Hermitian matrix $\boldsymbol \Sigma$ so that $\boldsymbol \Sigma (\mathbf L
  -x)\boldsymbol \Sigma =- (\mathbf L -x) $. As a result, the spectrum is
  symmetric about $x$. Here, we identify $\boldsymbol \Sigma=\boldsymbol
  \sigma_{\text z}$ and $x= -2 \overline \gamma$.}\BibitemShut {Stop}%
\end{thebibliography}

\newpage

\newpage

\widetext

\begin{center}
\textbf{\huge Supplementary information}
\end{center}


\newcommand{\ket}[1]{\left|#1\right>}
\newcommand{\nn}{\nonumber\\}
\newcommand{\f}[1]{\mbox{\boldmath$#1$}}

\section{Derivation of the rate equation}

In this section, we show how to derive Eq.~(2) in the letter starting from  the general Markovian master equation for a single quantum dot between two leads in the regime of high Coulomb interactions, which restricts   the Hilbert space to only zero or one electrons. 

Let  $d(d^{\dagger})$ denote the annihilation (creation) operator   of the quantum dot, and $\tilde \Gamma_{\nu}=\tilde \Gamma_{\nu}(\epsilon_{\text{dot}}) $   the spectral tunnel density into the corresponding reservoir $\nu= \rm R, L$ evaluated at the dot level $\epsilon_{\text{dot}}$. Using this notation, the master equation can be written in the form
\begin{equation}
\dot{\rho}  = \frac{\tilde \Gamma_{\text{out }  }}{2} \left(2d\rho d^{\dagger}-d^{\dagger}d\rho-\rho d^{\dagger}d\right)+ \frac{\tilde \Gamma_{\text{in}  }}{2}\left(2d^{\dagger}\rho d-dd^{\dagger}\rho-\rho dd^{\dagger}\right),
\end{equation}
with
\begin{equation}
\Gamma_{\text{out }  }=\tilde \Gamma_{\rm R}(1-f_{ \rm R})+\tilde \Gamma_{\rm L}(1-f_{\rm  L});\quad 
\Gamma_{\text{in} }=\tilde \Gamma_{\rm R}f_{ \rm R}+\tilde \Gamma_{ \rm L}f_{\rm L},
\end{equation}
where $f_{\nu}= f_{\nu} (\epsilon_{\text{dot}})$ is the Fermi function 
\begin{equation}
f_{\nu}(\epsilon)=\frac{1}{e^{(\epsilon-\mu_{\nu})/(k_{\text B} T_\nu)}+1} \ . \label{eq:fermi}
\end{equation}
Here, $\mu _{\nu}$ is the chemical potential of the lead $\nu$, $k_{\text B}$ is the Boltzman constant and $T_\nu$ is the temperature.
The two rate equations for the occupation $d=0,1$ can be written as  a $2\times2$ matrix:
\begin{equation}
\dot{\underline {\mathcal R } }={\cal L}\underline {\mathcal R } =
\begin{pmatrix}
-\tilde \Gamma_{\text{out }  } &\tilde \Gamma_{\text{in} }\\
\tilde \Gamma_{\text{out }   }  & -\tilde \Gamma_{\text{in} }
\end{pmatrix}
\underline {\mathcal R } \ ,
\label{eq:rateEquationSimple}
\end{equation}
where $\underline {\mathcal R } = \left(\tilde p_{1},\tilde p_{0} \right)^{\text T}$ is the vector of the corresponding probabilities $\tilde p_d$.

By extending the space, we can use the resolved probabilities $p_{m,d}$ of being in state $p_d$ while $m$ particles have been
counted tunneling out of the right reservoir.
According to Eq.~\eqref{eq:rateEquationSimple}, the rates for the relevant processes read\\[0.25cm]

\begin{tabular}{c|c|c}
initial state & final state & rate\\
\hline
$\ket{m,0}$ & $\ket{m,1}$ & $\tilde \Gamma_L f_L$\\
$\ket{m,0}$ & $\ket{m+1,1}$ & $\tilde \Gamma_R f_R$\\
$\ket{m,1}$ & $\ket{m,0}$ & $\tilde \Gamma_L (1-f_L)$\\
$\ket{m,1}$ & $\ket{m-1,0}$ & $\tilde \Gamma_R (1-f_R)$
\end{tabular}\\[0.25cm]

We can also write this as differential equations, where
\begin{align}
\dot{p}_{m,1} &= \tilde \Gamma_L f_{\text L} p_{m,0} + \tilde \Gamma_{\text R} f_{\text R} p_{m-1,0} - \left(\tilde \Gamma_{\text L} (1-f_{\text L})+\Gamma_{\text R}(1-f_{\text R})\right) p_{m,1}\,,\nn
\dot{p}_{m,0} &= \tilde \Gamma_{\text L} (1-f_{\text L}) p_{m,1} + \tilde \Gamma_{\text R}(1-f_{\text R}) p_{m+1,1} - \left(\tilde \Gamma_{\text L} f_{\text L} + \tilde \Gamma_{\text R} f_{\text R}\right) p_{m,0}
\,.
\end{align}
Alternatively, organizing the probabilities in a large vector, we get
\begin{align}
\frac{d}{dt} 
\left(\begin{array}{c}
\vdots\\
p_{m,1}\\
p_{m,0}\\
\vdots
\end{array}
\right)
=
 \underline {\dot \rho'}
=
\left(\begin{array}{ccccc}
\ddots & \ddots & \ddots\\
 & \mathbf L_+ & \mathbf  L_0 & \mathbf  L_- &\\
 & & \ddots & \ddots & \ddots
\end{array}\right)
\left(\begin{array}{c}
\vdots\\
p_{m,1}\\
p_{m,0}\\
\vdots
\end{array}\right)
= \mathbf L' \underline \rho'
\,,
\label{eq:rateEquationVector}
\end{align}
where the $\mathbf  L_x$ are $2\times 2$ matrices of the form
\begin{align}
\mathbf  L_0 &= \left(\begin{array}{cc}
 -\tilde \Gamma_{\rm  L}(1-f_{\rm  L} )-\tilde \Gamma_{ \rm R}(1-f_{\rm  R})  & +\tilde \Gamma_{\rm  L} f_{ \rm L}   \\
+\tilde \Gamma_{\rm  L} (1-f_{\rm  L})  &  -\tilde \Gamma_{\rm  L} f_{ \rm L} -\tilde \Gamma_{\rm  R} f_{\rm  R}
\end{array}\right)\,,\nn
\mathbf  L_- &= \left(\begin{array}{cc}
0 & 0\\
+\tilde \Gamma_{\rm  R} (1-f_{\rm R}) & 0
\end{array}\right)\,,\qquad
\mathbf  L_+ = \left(\begin{array}{cc}
0 & +\tilde \Gamma_{\rm  R} f_{\rm R} \\
0 & 0
\end{array}\right)\,.
\end{align}
We see that the entries of each column of $\mathbf  L$ matrix all add up to zero, which guarantees probability conservation. 	 
Demanding that the matrix $\mathbf  L$ is hermitian, we get the equations
\begin{align}
\tilde \Gamma_{\rm L} f_{\rm L} = \tilde \Gamma_{ \rm L} (1-f_{ \rm L})\,,\qquad
\tilde \Gamma_{\rm R} f_{ \rm R} = \tilde \Gamma_{ \rm R} (1-f_{\rm  R})\,,
\end{align}
which can only be fulfilled by $f_{ \rm L}=f_{ \rm R}=1/2$. 
This condition  is achievable at equilibrium $\mu_{\rm  L} = \mu_{\rm R} =\epsilon_{\rm dot}$ or close to equilibrium for a sufficiently large temperature such that $\mu_{\nu}-\epsilon_{\rm dot} \ll k_{\rm B}  T$.

Finally, for a notational reason we define $\Gamma_\nu \equiv \frac 12 \tilde \Gamma_\nu$, so that Eq.~\eqref{eq:rateEquationVector} becomes equivalent to Eq.~(2) in the letter. The effective coupling parameters are thus proportional to the spectral coupling density.

\section{Escape time statistics for a finite-size SSH model}

Here we provide detailed information about the derivation of Eq.~(5) in the letter, which gives an analytic expression for the escape time distribution. The derivation is adapted from Ref.~\cite{Brandes2008} and is adjusted for our purposes.

In order to derive Eq.~(5) in the letter, we consider a representation of the equations of motion Eq.~(2) in the letter using an infinite dimensional vector space, where the probabilities are the elements of the probability vector
\begin{equation}
\underline \rho'=
\left(
\begin{array}{c}
	\vdots\\
	p_{0}\\
	p_{1}\\
	\vdots\\
	p_{N}\\
	\vdots
\end{array}
\right)=
\left(
\begin{array}{c}
	\vdots\\
	\underline \rho\\
	\vdots
\end{array}
\right),
\end{equation}
where $\underline \rho$ is the vector containing the probabilities corresponding to the SSH section of the chain as used in Eq.~(4) in the letter. In order to distinguish the finite-size vector space of the SSH model and the infinite vector space under consideration here, we label the  vector in the latter vector space with $\underline \rho'$.
The Liouvillian in Eq.~\eqref{eq:rateEquationVector} for $f_\nu  = \frac 12$ becomes
\begin{equation}
 \mathbf L' = \left(\begin{array}{cccccc}
\ddots & \ddots\\
\ddots & \ddots & \Gamma_L\\
& \Gamma_{\rm L} & -\Gamma_{\rm L}-\Gamma_{\rm R} & \Gamma_{\rm R}\\
& & \Gamma_{\rm R} & \ddots & \ddots\\
& & & \ddots & \ddots & 
\end{array}\right)\,,
\end{equation}
The Liouvillian $ \mathbf L'$ can be written as
\begin{equation}
\mathbf L' = \mathbf L_0' + \mathbf J',
\end{equation}
where $\mathbf L_0'$ has a block-diagonal structure 
\begin{equation}
 \mathbf L_0' = 
 \left(
 \begin{array}{ccc}
  \mathbf L_{\rm t} &   & \\
   &  \mathbf L_{\rm SSH} & \\
  & &  \mathbf L_{\rm b}
\end{array}
\right)
.
\end{equation}
Thereby, $\mathbf L_{\rm SSH}$ is the block with dimension $N\times N$ referring to the middle section of the infinite chain corresponding to the sites $n=1,\dots,N$. The matrices $ \mathbf L_{\rm t}$ and  $ \mathbf L_{\rm b}$ denote infinite-dimensional matrices describing the dynamics in the top and bottom sections of the infinite chain, respectively. The infinite-dimensional jump matrix $\mathbf J$ contains four entries. They read
\begin{equation}
	\left(\mathbf J \right)_{ij} = 
	\left( \delta_{i,1} \delta_{j,0}+  \delta_{i,0} \delta_{j,1} \right)\Gamma_R+ \left( \delta_{i,N} \delta_{j,N+1}+  \delta_{i,N+1} \delta_{j,N} \right)\left[\overline \gamma-(-1)^N\alpha \right],
\end{equation}
and connect thus the blocks appearing in $\mathbf L'$. 

In an interaction picture with respect to $\mathbf L_0$, the solution of Eq.~\eqref{eq:rateEquationVector} reads
\begin{equation}
\underline{\rho}'(t)= \sum_{k=0}^{\infty} \int_{0}^{t} dt_k \dots \int_{0}^{t_2} dt_1 \underline \rho_{t,k}\left( \left\lbrace t_i\right\rbrace\right),
\end{equation}
where
\begin{equation}
\underline \rho_{t,k}\left( \left\lbrace t_i\right\rbrace\right)  =
e^{\mathbf L_0'(t-t_k)} \mathbf J' e^{\mathbf L_0'(t_k-t_{k-1})}  \mathbf J'\dots \mathbf J' e^{\mathbf L_0't_1} \rho_{}'(0)
\end{equation}
is a conditioned state of the system, where the unperturbed time evolution corresponding to $\mathbf L_0'$ has been interrupted by the processes corresponding to $\mathbf J'$ at times $t_i$ with $i=1,..,k$. In particular, $\underline \rho_{t,k=0}$ describes a time evolution with no jumps within the time interval $t'\in (0,t)$. For this reason, we interpret 
\begin{equation}
	P(t) = \underline u^{\rm T} \mathbf J' e^{\mathbf L_0 t}\rho_{}'(0) 
\end{equation}
as the probability distribution that the process described by $\mathbf J'$ takes place at time $t$ for the first time. Thereby, we have defined the infinite dimensional vector $\underline u^{\rm T}=(\dots,1,1,1,1,\dots)$.

In doing so, we find for the integrated probability distribution
\begin{align}
	\int_{0}^{\infty} P(t)dt &=- \underline u^{\rm T} \mathbf J' \mathbf L_0^{-1}\rho_{}'(0) \\
	         &= -\underline u^{\rm T} \left( \mathbf L' -\mathbf L_0 \right)  \mathbf L_0^{-1}\rho_{}'(0)
	         = \underline u^{\rm T}\rho'(0) =1.
\end{align}
Moreover, one can show that $P(t)>0$~\cite{Brandes2008}, so that $P(t)$ indeed defines a probability distribution. 

Assuming an initial condition $\underline \rho'(0)$ restricted to the sites of the SSH chain, we find
\begin{align}
	 P_{\rm e}(t) 
	= \underline J^{\rm T} e^{\mathbf L_{\rm SSH} t} \rho(0),
\end{align}
where $\underline J\in IR^N$ is given by the elements of the vector $\underline J'=  \mathbf J' \underline u$, which correspond to the SSH section of the vector space. This is  thus the expression for the escape time distribution given in Eq.~(5) in the letter. Thereby, we have used that the dynamics determined by the block-diagonal matrix $\mathbf L_0$ does not allow the system to escape from SSH section. The vector $\underline J$ is equivalent to the jump vector defined in the letter.

\section{Generalized inversion symmetry for an odd number of sites in the SSH model}

The finite-size SSH system with an open boundary condition and an odd number of sites $N$ fulfills a generalized inversion symmetry
\begin{equation}
\mathbf P  \mathbf L_{\rm SSH} \mathbf P= \mathbf  L_{\rm SSH},
\end{equation}
where
\begin{equation}
\mathbf P =
	\left(
	\begin{array}{cccccccccccc}
	a  &  0& a r& 0  &  a r^2& 0&\dots  & a r^{(N-5)/2} & 0& a r^{(N-3)/2} &  0 & b  \\
	0  &  0 & 0  &   0&  0&\dots&    \dots & 0& 0 & 0             & 1   & 0   \\  
	 a r & 0  &  a r^2  & 0 &\dots&  \dots & \dots & a r^{(N-3)/2} &  0&  b  & 0 &  s a              \\
	 0  &  0& 0 &\dots&   \dots  &  \dots  & \dots  &   0   & 1    &    0    & 0   & 0  \\
	a r^2  & 0&\dots& \dots &  \dots  & \dots \dots & \dots & b  &  0&  sa    & 0 &  s a r              \\
	0 &\dots &  \dots & \dots   &  \dots &  \dots & \dots &   0   & 0    &   0     & 0   & 0  \\
	\vdots &\vdots & \vdots& \vdots &   \vdots&  \vdots&  &   \vdots   & \vdots    &         & \vdots  & \vdots  \\
	a r^{(N-5)/2}   &  0 & a r^{(N-3)/2}   &   0& b &0&    \dots &  s a r^{(N-11)/2}& 0 &  s a r^{(N-9)/2}             & 0   &  s a r^{(N-7)/2}   \\  
	0  &  0 & 0  &   1& 0 & 0 &    \dots & 0& 0 & 0             & 0   & 0   \\  
	a r^{(N-3)/2} & 0 & b  &  0& s a & 0  &  \dots& s a r^{(N-9)/2}& 0 &  s a r^{(N-7)/2} & 0 & s a r^{(N-5)/2} \\
	0  &  1 & 0  &   0 & 0& 0 &    \dots & 0& 0 & 0             & 0   & 0   \\  
	b & 0 & s a  &  0& s a r& 0  & \dots & s a r^{(N-7)/2} & 0 & s a r^{(N-5)/2} & 0 & s a r^{(N-3)/2} \\
	\end{array}
	\right).
\end{equation}
For a notational reason we  have introduced $r=-\Gamma _{\text L}/ \Gamma_{\text R}$ and
\begin{align}
b&= r\left(   r^{(N-3)/2}a-1 \right)  ,  \nonumber \\
a&= \frac{1}{\Sigma + r^{N-1}} \left( 2 \left| r^{(N-3)/2}  \right| r + \sqrt{   1-r^N + r^{N-1}} \right)(-1)^{(N-1)/2},\nonumber \\
\Sigma&= \frac{1-r^N}{1-r^2},\nonumber \\
s &=(-1)^{(N-1)/2}.
\end{align} 
The matrix $\mathbf P $ is Hermitian and unitary. Surprisingly, it decouples even and odd sites $n$. When restricting the matrix to the even-sites subspace, then we find that the corresponding matrix is equivalent to a usual inversion matrix 
\begin{equation}
\mathbf P_{\rm even} =
	\left(
	\begin{array}{ccccccccc}
	0   & 0 & 0& 0&\dots& 0& 0 & 0& 0\\
	0   & 0 & 0 & 0 &\dots& 0& 0 & 1& 0\\
	0 & 0 & 0 & 0&\dots & 0& 0& 0& 0\\
	0 & 0  & 0 & 0&\dots& 1& 0& 0& 0\\
	\vdots   & \vdots & \vdots& \vdots &\vdots &\vdots  &\vdots& \vdots& \vdots\\
	0 & 0 & 0 & 1&\dots & 0& 0& 0& 0\\
	0 & 0  & 0 & 0&\dots& 0& 0& 0& 0\\
	0 & 1 & 0 & 0&\dots & 0& 0& 0& 0\\
	0 & 0  & 0 & 0&\dots& 0& 0& 0& 0
	\end{array}
	\right).
\end{equation}
For this reason, any eigenstate $\underline v_j$ of $ \mathbf L_{\rm SSH} $ fulfills a symmetry relation $v_{j,n}=\pm v_{j,N+1-n} $ for $n$ even. Consequently, all eigenstates have an even or odd symmetry under this generalized inversion symmetry.

Using this property, we can now explain, why half of the coefficients $a_j $ in Fig.~2(c) in the letter vanish. Using the time-independent Schr\"odinger equation with $\mathbf L_{\rm SSH}$ and  $v_{j,n}=- v_{j,N+1-n} $ for an odd eigenstate and $n=2$, it is straight forward to show that
\begin{equation}
a_j = \underline v_{j}\cdot \underline J =\Gamma_{\text  R } v_{j,1} +  \Gamma_{\text  L } v_{j,N} = 0.
\end{equation}

\section{Construction of the waiting time distribution}

\label{app:waitingTimeConstruction}

In this section we explain step by step how to obtain the reconstructed waiting time distribution (WTD) and the integrated WTD from a set of escape times $\left\lbrace t_{\text{e},i}\right\rbrace$.

First, we sort the set of  escape times so that $t_{\text{e},i}<t_{\text{e},j}$ for $i<j$. Next, we construct a set of  pairs  $\left\lbrace\left\lbrace     t_{\text{e},i} ,i \right\rbrace \right\rbrace $ and apply a moving average so that we obtain the smoothened set
$\left\lbrace \left\lbrace \overline t_{\text{e},i} , \mathcal A _i\right\rbrace \right\rbrace $ with
\begin{align}
\overline t_{\text{e},i}  &= \frac{1}{N_{\rm av} }  \sum_{j=i}^{i+N_{\rm av}}  t_{\text{e},j}  ,\nonumber \\
	\mathcal A _i  &= \frac{1}{N_{\rm av} } \sum_{j=i}^{i+N_{av}} j .
\end{align}
In our calculations we take $N_{\rm av}=100$. In order to enable an efficient numerical fitting algorithm, we take only a subset $\mathcal I$ of $\left\lbrace\left\lbrace\overline t_{\text{e},i}   \mathcal A _i  \right\rbrace \right\rbrace $. More precisely, we choose $\mathcal I=  \left\lbrace\left\lbrace  \overline t_{\text{e},i}  ,\mathcal A _i \right\rbrace\vert i= N_{\rm step}\cdot l \; \text{and} \; l=0,1,2,\dots \right\rbrace $ with $N_{\rm step}=500$.

This set $\mathcal I$ is then the reconstructed integrated WTD as depicted in Fig.~3 (a) in the letter, which we take for the fitting procedure with results depicted in the bottom row of Fig.~2 in the letter.
The reconstructed waiting time distribution is  given by the set
\begin{equation}
\mathcal W=  \left\lbrace\left\lbrace \mathcal \overline t_{\text{e},i}  , \frac{\mathcal A_{i+N_{\rm st}}- \mathcal A_{i} }{\overline t_{\text{e},i+N_{st}}-\overline t_{\text{e},i}} \right\rbrace\vert i= N_{\rm step}\cdot  l\; \text{and} \; l=0,1,2,\dots \right\rbrace.
\end{equation}

\section{Dependence on the number of fit terms $K$ and the number of experimental runs}

\label{app:fitTermsExperimentalRuns}

\begin{figure}[t]
\centerline{\includegraphics[width=0.5\linewidth]{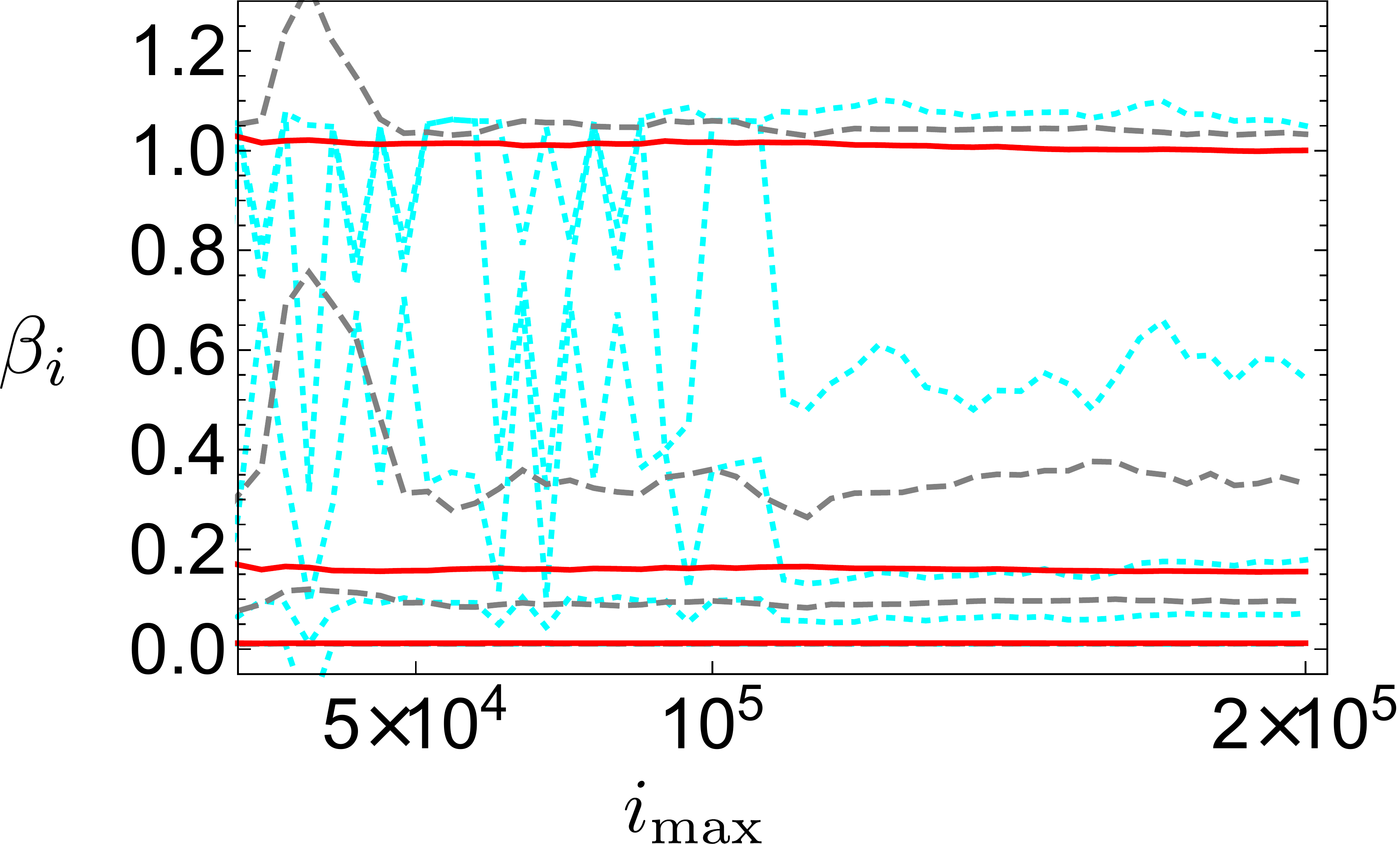}}
\caption{ Dependence of the exponent spectrum as a function of the number of runs $i_{\rm max}$ used to construct the integrated WTD. Overall parameters are as in Fig.~3 in the letter. For the red (solid), gray (dashed)  and cyan (dotted) curves we used $K=3$, $K=4$ and $K=5$, respectively.
}
\label{fig:otherInformation}
\end{figure}%
%

Here, we discuss the performance of the fitting procedure as a function of the number of stochastic trajectories $i_{\rm max}$ used to reconstruct the WTD. In Fig.~\ref{fig:otherInformation}, we depict the spectrum of exponents $\left\lbrace \beta_j\right\rbrace$ as a function of different number of experimental runs $i_{\rm max}$ for different number of fit terms $K$ in Eq.~(8) in the letter. We observe that for $\alpha=-0.5\overline \gamma $ and $K=3$ the exponents $\beta_j$ exhibit a stable value for a rather small number of experimental runs $i_{\rm max}=10^4$.  For $K=4$ and $K=5$ we find that the fitting procedure exhibits variations depending on the number of runs. We observe, that these variations continue even for higher number of runs. This effect appears as there are too many degrees of freedom  $\left\lbrace A_i\right\rbrace$ and $\left\lbrace\beta_i\right\rbrace$ in the fit procedure, so that there are several  combinations which exhibit a good fit quality. This results in  the variations visible in Fig.~\ref{fig:otherInformation}.

\section{Rates for tunnel-based feedback}

In the letter, we have focused on the simplest topological model that can be generated from the transport dynamics of the SET, the SSH model.
In this section, we explain how to modify and create more general random walks using feedback control in the context of a single-electron transistor. 
To this end,  directly after each measurement we adapt the tunneling rates $\tilde \Gamma_\nu \to \tilde  \Gamma_\nu^{n}$, according to the present state  $\left|n=2m-1\right>$ of the system~\cite{Schaller2014}. Consequently, the rate matrix in equation Eq.~\eqref{eq:rateEquationVector} now reads
\begin{align}
\mathbf L = \left(\begin{array}{ccccc}
\ddots &\ddots    & \\
\ddots & \mathbf L_0^{n-2} &\mathbf  L_-^n\\
& \mathbf L_+^{n-2} &\mathbf  L_0^n &\mathbf  L_-^{n+2}\\
& &\mathbf  L_+^n &\mathbf L_0^{n+2} & \ddots\\
& & & \ddots & \ddots
\end{array}\right)\,,
\end{align}
where the submatrices are  written as 
\begin{align}
\mathbf L_0^n &= \left(\begin{array}{cc}
-\tilde  \Gamma_{\rm L}^n f_{\rm L} -\tilde  \Gamma_{\rm  R}^{n} f_{\rm R} & +\tilde  \Gamma_{\rm L}^{n+1} (1-f_{\rm  L} )\\
+\tilde  \Gamma_{\rm L}^n f_{\rm  L}  & -\tilde  \Gamma_{\rm L}^{n+1} (1-f_{\rm  L})-\tilde  \Gamma_{\rm R}^{n+1} (1-f_{\rm R} )
\end{array}\right)\,,\nn
\mathbf L_-^n &= \left(\begin{array}{cc}
0 & 0\\
\tilde  \Gamma_R^n f_{ \rm R}  & 0
\end{array}\right)\,,\qquad
\mathbf L_+^n = \left(\begin{array}{cc}
0 & \tilde  \Gamma_{\rm R} ^{n+1}  (1-f_{\rm R})\\
0 & 0
\end{array}\right)\,.
\end{align}
Still, the entries in the columns of $\mathbf L$   add up to zero, which ensures the probability conservation. 	 

\begin{figure}[t]
\centerline{\includegraphics[width=0.9\linewidth]{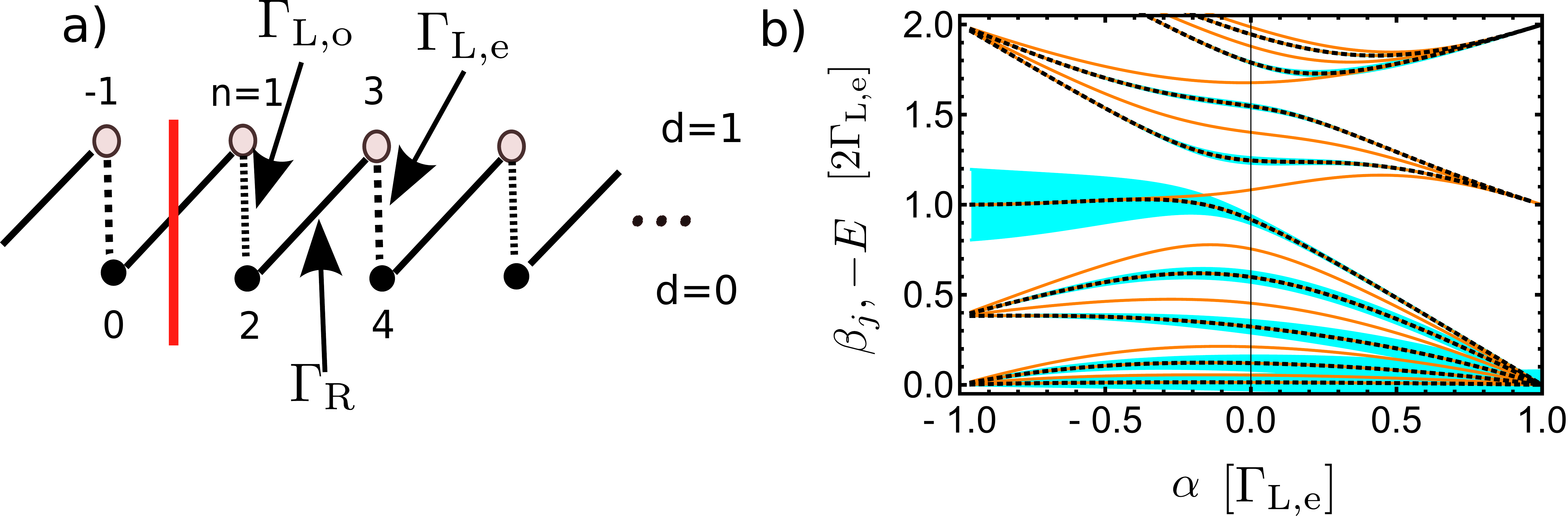}}
\caption{
(a) Sketch of a random walk which could be generated in a SET using feedback control. 
(b) depicts the same as Fig.~(2) in the letter but for this generalized model for $N=18$. Here we parameterize $\Gamma_{\rm R}=\Gamma_{\rm L,\text e}+\alpha$ and $\Gamma_{\rm L,\text o}=\Gamma_{\rm L,\text e}-\alpha$.
}
\label{fig:example}
\end{figure}

An example is depicted  in Fig.~\ref{fig:example}. It describes a random walk on the chain, where the rates repeat after four steps, thus $\Gamma_{\rm  R},\Gamma_{\rm L,\text  e},\Gamma_{ \rm R},\Gamma_{\rm  L,\text  o} $. Parameterizing the site $n=2m-d$, the tunnel rates shall read
\begin{align}
	\tilde \Gamma_{\rm  R} ^n &= \tilde \Gamma_{\rm  R} \nonumber, \\
	\tilde \Gamma_{\rm  L} ^{n=2m-d} &=   \tilde \Gamma_{\rm  L,0}  + (-1)^m \tilde \Gamma_{\rm  L ,1} .
\end{align}
 Thus, we condition the left tunnel rate on the number of particles $m$ which have been tunneled out of the right reservoir. More precisely, the left tunnel rate depends on $m$ odd and $m$ even. Additionally, we assume $f_\nu =\frac 12$ as in the letter. In doing so, $\mathbf L$ becomes Hermitian. The effective rates are thus
\begin{align}
\Gamma_{\rm R} &= \frac 12 \tilde \Gamma_{\rm R} ^n,\qquad
\Gamma_{\rm L,\text e}= \frac 12 \tilde \Gamma_{ \rm L} ^{n=2m-d} \quad \text{if } m  \text{ even },\qquad 
\Gamma_{\rm  L,\text o}= \frac 12 \tilde \Gamma_{\rm L} ^{n=2m-d} \quad \text{if } m  \text{ odd }.
\end{align}
Due to an inversion symmetry, this system can also exhibits a topological phase transition with corresponding midgap  modes (Fig.~\ref{fig:example}(b)).


\end{document}